# Aspetti moderni della fisica greca
*(Modern Aspects of ancient Greek Physics)*

Erasmo Recami
(*Facoltà di Ingegneria, Università statale di Bergamo, Bergamo, Italy*,
and *INFN-Sezione di Milano, Milan, Italy*)

**Abstract in English:**  The translations into modern languages of the books and papers by ancient scientists (let us confine ourselves, here, to physicists) have been the praiseworthy work of scholars whose fields of competence, however, were too often in humanities, rather than in science.  Such translations, therefore, miss points important for modern physics, since for a good translation it is of course necessary a good understanding of the topic. [Even in the case of a rather recent author like Leonardo da Vinci, whose language was not that different from modern Italian, many translations into current languages are misleading]. In our opinion, scientists with a deep knowledge in physics, and a rare interest in its history, have to revise all the enourmous body of the ancient physicists' writings. To give an idea of the terrible effort that should be undertaken in this respect by physicists, we call attention in this brief article to some extremely "modern" results, frequently overlooked, got by ancient Greek physics. Since the present e-print is in Italian, we bold-faced some of the most interesting statements in order to reduce the extension of the material to be possibly examined by Italian-ignoring people.  Let us moreover translate here, into English, just two remarkable claims:  (i) Posidonius (circa135-51 BC) wrote: *«Matter is endowed with a cohesion that keeps it together and against which the surrounding vacuum has no power. Indeed, the material world is supported by an immense force, and alternately contracts and expands in the vacuum following its own physical processes, now consumed by fire, now, instead, giving rise to a new creation of the cosmos»*;  (ii) Plutarchus, circa 100 AD, in his early book on "astrophysics" --in which he exposed, in a sense, a general theory of gravitation-- wrote the noticeable passage: *«The Moon gets the guarantee of not falling down just from its motion and from the dash associated with its revolution, exactly as stones in slings cannot fall due to their circular whirling motion; in fact, each thing is dragged by its mere natural motion only if it isn't deviated by something else. The Moon, therefore, is not dragged down by its weight, because its natural tendency is frustrated by its revolution.  And, on the contrary, it would be really amazing if it remained  at rest always at the same place, like the Earth».*

# 1. - Introduzione.

Ogni "storia della fisica", in senso moderno, è da intendersi soprattutto come una *storia del pensiero* scientifico-fisico; pensiero il quale ai giorni nostri ha al suo attivo contenuti culturali, formativi, filosofici, logici, che ben si possono dire il prodotto caratteristico e l'eredità del XX secolo. Se il fine della cultura (certo non risolvendosi nel possesso di nozioni) sta nel fornire una immagine possibilmente unitaria del mondo, risulta infatti necessario considerare anche l'orientamento che in questo senso ci proviene dalla scienza, ed in particolare dalla fisica.

Di più: la cultura consiste nell'acquisto di una visione globale della vita e del mondo *in cui si vive,* e nel possesso dei mezzi per un giudizio critico sugli avvenimenti *che ci circondano.* A maggior ragione, perciò, proprio nel nostro tempo --in cui tanto si invoca giustamente un nuovo umanesimo scientifico, e, soprattutto, in cui la vita subisce profondamente l'influenza della tecnica, sorella della fisica e delle scienze-- sarebbe opportuno che i fondamenti della fisica moderna fossero, in modo preciso e maturo, a conoscenza del più. vasto pubblico od almeno delle persone colte *e dei giovani.* D'altra parte, col progredire della specializzazione delle ricerche teoriche e sperimentali su oggetti lontanissimi dall'esperienza quotidiana, e del conseguente uso di un linguaggio matematico via via più astratto e complesso, la fisica tende effettivamente a diventare di per sé un mondo sempre più lontano da quello comune. Questa difficoltà può essere però in parte "didatticamente" elusa col limitarsi ai fondamenti "filosofici" della fisica recente, comprensibili *a tutti* --*a priori*-- nei loro aspetti "umanistici".

A tale scopo è forse utile rifare passare davanti ai nostri occhi le varie posizioni assunte nel tempo dai fisici di fronte alla natura. Si cercherà, in ciò, di dare il maggior risalto alle anticipazioni della scienza recente; anche se non si potrà parlare della maggior parte della fisica *contemporanea,* e per la sua enorme estensione e per la difficoltà di fare la storia di dottrine tuttora in fase di travaglio. Così l'esame della fisica greca ci fornirà lo spunto per ricordare vari aspetti della fisica recente.

In secondo luogo si trascurerà la "cronaca" storica in confronto allo sviluppo dei *fondamenti* della fisica. Ma non sembri strano che si accenni di sfuggita a molti scienziati che meriterebbero attenzione. È purtroppo ine-vitabile che rimangano per lo più addirittura sconosciuti gli innumeri fisici i quali spesso dedicarono la propria vita alla ricerca, non raramente a costo di molti sacrifici. Preferiamo ricordare *idealmente* questa grande schiera di uomini (dai tempi preistorici a quelli, in particolare, degli ultimi quattro secoli), il cui lavoro non fu superfluo.

La collaborazione di questi moltissimi sconosciuti, con la loro naturale varietà di interessi e talenti, fu necessaria per produrre la messe di osservazioni e calcoli che permise i progressi successivi. L'attività di costoro, studiosi ora dimenticati, costituì e continua a formare l'indispensabile base di partenza per ogni importante conquista, compresa quella del genio. Dalla fine del diciassettesimo secolo, la fisica è stata una costruzione fondata sulla cooperazione di molti: anche questo è un fatto storico.

Bisogna, poi, aggiungere che la stessa questione della priorità scientifica costituisce quasi sempre un argomento sfortunato: anche oggidì la ricompensa della fama viene solo a chi ha buona sorte (cioè risolve un problema nel momento giusto e pubblicando i risultati nel "posto" giusto) oppure ha la capacità di *diffondere* ampiamente per primo (non di *avere* per primo) una nuova idea. Inoltre la storia ricorda soprattutto gli esperimenti che hanno avuto successo e le teorie che sopravvivono, almeno per un certo tempo. Ciò può dare la falsa impressione che la storia della fisica sia una serie di trionfi; ma non si deve dimenticare che per ogni successo ci furono molti tentativi falliti che gli fanno da "sfondo storico", e che soli potrebbero illustrare l'originalità che vi occorse.

Infine, non ci stupiranno le parole del Premio Nobel Kastler: «La scienza, come l'arte, è un'opera umana in pieno sviluppo ed a carattere dinamico». Anche la fisica --moderna e libera creazione dello spirito umano, benché venga evolvendosi con sue caratteristiche proprie-- richiede in ogni caso doti di intuizione, creatività, fantasia, senso estetico (non meno che intelligenza, logica, spirito critico e di osservazione quantitativa, onestà, precisione, costanza, abilità tecnica in senso sperimentale o teorico-matematico, ecc.). Quanto poi essa sia dinamica appare non solo dall'esperienza contemporanea, bensì anche dal suo passato. Si noti per esempio quanto mutò la posizione degli scienziati di fronte al conoscibile, come si può arguire dai passi di Kepler, Newton, Eddington ed Heisenberg qui di séguito citati. «Poiché io mi sono assunto il compito di permettere all'intelletto umano, con l'ausilio del calcolo geometrico, di gettare uno sguardo sulle vie della Sua creazione, voglia l'Artefice dei cieli, il Padre immortale di tutti gli esseri intelligenti, al quale tutti i sensi mortali debbono la loro esistenza, essermi clemente e guardarmi dal dir sulla Sua opera cose che non possano conciliarsi colla Sua maestà ed inducano in errore il nostro intelletto, e far sì che noi imitiamo la perfezione della Sua opera creatrice santificando la nostra vita». Tale la professione di fede di Kepler, che palesemente si considerava sul punto di contemplare il piano creativo di Dio e di inchinarsi riverente di fronte al santuario così svelato. Contemporaneamente si affermava il fine della fisica: determinare nel manifesto divenire dell'universo, al di sotto delle apparenze e delle "forme geometriche", ciò che rimane costante, ossia la legge del divenire stesso. Ma si sa che le conquiste della fisica, con l'approfondirsi della conoscenza che essa ha di sé (oltre che per la conoscenza dell'uomo che essa dà all'uomo), presentano la caratteristica di comportare parallelamente una serie sempre maggiore di rinunce; e già Newton --solo una cinquantina d'anni dopo-- doveva scrivere, come ben noto: «Io non so come mi giudica il mondo; a me sembra di essere un bambino che gioca sulla spiaggia del mare e si rallegra se di quando in quando trova un ciottolo più liscio degli altri od una conchiglia più bella delle altre, mentre il grande oceano della verità sta inesplorato dinanzi a lui». Già con Newton lo scienziato è, dunque, soltanto alla porta di una terra nuova ed infinita, di cui non può scorgere in nessun punto i confini, ed il suo compito è ora di analizzare con precisione non tanto *la* legge universale, quanto molte singole leggi *particolari*. Tuttavia restava la certezza di conoscere il vero comportamento delle cose, il cui mutamento si compiva mediante il loro moto nello spazio (anche il mutare delle qualità essendo riconducibile, si sa, al movimento delle parti più piccole). *Dopo* oltre due secoli di evoluzione, un clima ben diverso guida la penna del Premio Nobel Heisenberg: «Noi ci rendiamo conto... che non c'è un sicuro punto di partenza per le vie che conducono ai vari campi dello scibile, ma che ogni conoscenza è sospesa in certo modo sopra un abisso senza fondo; che dobbiamo sempre cominciare da qualche "punto di mezzo", mentre gli stessi termini usati per parlare dei fenomeni acquistano a poco a poco un senso più preciso solo col loro uso». Inoltre, nella fisica classica newtoniana, si era sempre trascurato l'apporto del *soggetto* nell'indagine, considerandolo passivo nei confronti dei fenomeni; l'osservatore si era dimenticato cioè di costituire, con l'oggetto di osservazione, un sistema fisico in interazione. Ecco invece cosa arrivò a scrivere, noi primi decenni del nostro secolo, il noto astronomo Eddington -- riferendosi alle "correlazioni d'incertezza" di Heisenberg--, con linguaggio gocciante retorica ma immaginoso: «Abbiamo visto che là dove la scienza ha compiuto le massime conquiste, lo spirito ha riavuto dalla natura ciò che egli stesso le aveva prestato. Sui lidi dell'ignoto abbiamo scoperto un'orma misteriosa; abbiamo escogitato... profonde teorie per riuscire a scrutarne l'origine. Alfine siamo riusciti a ricostruire l'essere da cui quella orma deriva: quell'essere siamo noi stessi». Ed ancora, riferendosi all'interpretazione statistico-probabilistica del valore di certa fisica recente, l'Eddington non si trattenne dal dire: «Io posso annunciare che la scienza che studia il mondo fisico ha voltato oramai le spalle a tutti siffatti modelli meccanici, che anzi sono considerati piuttosto come ostacoli alla comprensione di quella verità che dietro ai fenomeni sensibili si asconde».

Si vedranno, dunque, più da vicino le evoluzioni che condussero a tutti questi mutamenti prospettici, cominciando essenzialmente –con questo primo articolo di una serie di due-- dalla fisica antica e soprattutto dalla fisica greca.

2. - **La fisica pre-greca**.

Da quando l'uomo esiste come uomo, cioè diciamo da alcuni *milioni* di anni, egli svolse molto probabilmente attività religiosa, artistica, scientifica --che inizialmente fu in sostanza *tecnologica*--, seppellendo poi i morti, eseguendo riti magici, dipingendo caverne, preparando ed usando il fuoco, e strumenti, e così via. Caratteristica essenziale dell'umanità --nella scala evolutiva-- è proprio l'uso sistematico di attrezzi, intesi come prolungamenti degli organi e come alternativi ad un potenziamento delle proprie facoltà organiche. Di fronte ad ogni bisogno l'uomo si procurò lo strumento adatto, evitando ]a specializzazione del proprio organismo, a cui veniva lasciata intatta la *plasticità* necessaria per future evoluzioni e per il facile adattamento a nuove condizioni di vita. Ad esempio, l'adozione del randello e della selce scheggiata evitò l'esigenza di possedere forti mascelle, che avrebbe contrastato l'ampliamento della capacità cranica, e contemporaneamente comportò un perfezionamento delle mani --meraviglioso strumento, non piu schiavo del bisogno ma al servizio diretto della mente e dell'animo. (Anche i moderni *computer,* eseguendo la parte *meccanica* delle elaborazioni matematiche, tendono ad eliminare il

"manovalaggio" del cervello, lasciando questo libero di dedicarsi ad attività più elevate e più specificamente umane). I manufatti più antichi sembrano ascendere a circa due milioni d'anni fa.

È doveroso registrare anzitutto l'enorme sforzo tecnologico, fino a tempi a noi vicini, dell'umanità preistorica, alle prese con la costruzione di oggetti in pietra, osso, legno, creta, per la caccia, la pesca, la navigazione, il trasporto, l'abitazione, ecc. Questa costruzione richiese di necessità una vasta gamma di conoscenze empiriche
trasmissibili per insegnamento diretto -- pur non escludendo attività speculative. L'evoluzione delle tecniche di scheggiatura avvenne a prezzo di molte intelligenti scoperte, ed una vera invenzione --ad esempio-- portò poi all'arte fittile. D'altra parte si ricordi che i famosi dipinti e graffiti delle grotte dei Pirenei risalgono a varie decine di migliaia di anni fa, mentre altrove se ne sono trovati di molto più antichi. Ancora: circa dieci mila anni fa --per quanto finora noto-- i manufatti in ossidiana di Lipari venivano già regolarmente esportati, ad esempio in Liguria. L'evoluzione delle conoscenze fu lenta (e ci appare lentissima, per errore prospettico) anche per la mancanza di mezzi atti alla loro diffusione. Non si dimentichi che il veloce sviluppo della fisica e delle scienze di questi ultimi secoli *seguì* all'invenzione della stampa.

Ogni storia della fisica dovrebbe quindi iniziare con un esame della tecnologia preistorica; ma, per consuetudine, si rimanda ai competenti testi di paletnologia.

Saltate anche le ere dei metalli, passiamo all'Egitto. Gli Egizi, come i Babilonesi, raggiunsero i successi più notevoli nella scienza empirica dei movimenti stellari (compilarono anche calendari) e della matematica applicata. Le loro cognizioni riguardarono essenzialmente settori tecnici. L'alto livello della tecnologia egizia nell'arte mineraria (con sfruttamento di miniere a grande profondità), metallurgica, edilizia non fu superato per moltissimi secoli. Gli Egizi svilupparono nel corso dei millenni (uno dei primi re di cui si abbia precisa notizia storica, Menes, regnò dal 3200 a.C. circa) svariati sistemi per innalzare stupefacenti edifici in pietra, perfezionando contemporaneamente i metodi per risolvere i problemi subordinati, come estrazione dalle cave, trasporto e sollevamento di massi enormi, erezione di colonne e di obelischi giganteschi. Tutto ciò non sarebbe stato possibile senza larghe conoscenze pratiche di meccanica o di statica ed un buon livello tecnico dell'ingegneria, e richiese l'invenzione delle "macchine semplici" (basate sui principi fondamentali della meccanica): la leva ed il piano inclinato. Tra gli strumenti rinvenuti dagli archeologi ricordiamo squadre, livelle, fili a piombo, bilance a pesi, e (nei dipinti, almeno) telai da tessitore, navi a vela, carri a ruota trainati da cavalli, ecc. La tecnica egizia eclissa completamente quanto compiuto poi da Greci e Romani nel campo dell'edilizia. La cultura greca, in particolare, per eccessivo spregio delle attività pratiche (ad esempio quelle manuali erano eseguite spesso dagli schiavi) e per eccesso di mentalità aristocratica --che ha ancora le sue ultime funeste conseguenze in parte dell'Europa contemporanea-- non si dedicò mai alla tecnologia.

L'universo egizio era immaginato come una scatola rettangolare: l'Egitto ne occupava il centro della base, il cielo era sostenuto da quattro picchi montuosi, le stelle erano lampade sospese al cielo mediante funi. Intorno alla Terra scorreva un fiume, su cui navigava un battello che trasportava il Sole.

Anche in Babilonia, tra i cui primi abitanti storici ci furono i Sumeri, si ebbero vaste conoscenze astronomiche. Si pensa che già verso il 4000 a.C. venissero fatti tentativi per misurare il numero dei mesi nel ciclo delle stagioni; la conquista più importante fu però il riconoscimento dell'utilità di unità di misura fisse: verso il 2500 a.C. decreti reali stabilirono misure di lunghezza, peso e capacità. I **Sumeri** usarono i sistemi di numerazione decimale e duodecimale, ed il numero 60 ebbe perciò speciale rilievo (ancora lo usiamo nelle misure di tempo). Dal sesto secolo a.C. cominciarono a predire le eclissi. Ma a tutte queste conoscenze sempre continuarono a mescolarsi le concezioni magiche: per secoli il pensiero europeo (per non parlare di ora) fu influenzato dalle idee babilonesi sulle virtù di speciali numeri e sulla predizione del futuro mediante diagrammi geometrici.

Contemporaneamente a quelle egizia e babilonese si ebbero le civiltà della Cina e dell'India, che restarono isolate. Nel terzo secolo a.C. fu inventato lo schema dei numeri tuttora usati, cosiddetti "arabi". Venne formulata una teoria atomica ante litteram, che circa cento anni prima di Cristo fu estesa al *tempo,* inteso così come discontinuo (ogni oggetto è una serie di "esistenze" istantanee successive). Nel quarto secolo d.C. si ebbe notizia che la bussola era già usata da tempo dai Cinesi.

Accenniamo infine a Creta e Micene. Attraverso i passi montani di Creta esistono resti di strade gettate con tecnica avanzata; e nel palazzo reale si rinvennero opere di scienza applicata: meccanica, idraulica, ecc. Ma la scrittura cretese non è stata ancora decifrata. In quanto alla grande civiltà micenea, essa fu distrutta forse verso il 1400 a.C. da un cataclisma (così come Cnosso).

3. - **La fisica greca.**

Dall'insieme dei risultati --pur ammirevoli-- degli Egizi o dei Babilonesi non scaturisce un quadro uniforme, perché essi erano basati su conoscenze soprattutto pratiche, che non formavano un corpo unitario di pensiero scientifico.

Ciò si realizzò solo più tardi, quando si ebbe quel tentativo di razionalizzare i fenomeni (e spiegarli nel quadro di ipotesi generali), che costituì la grande creazione dei *Greci* del VI secolo a.C. Il Cosmo venne allora considerato come una singola unità ordinata, con leggi passibili di scoperta. Non è perciò casuale se i primi passi verso un pensiero sistematico furono intrapresi nell'àmbito della filosofia (ancora oggi --come è noto-- certe parti, assiomatizzate, della fisica vengono considerate come branche della "filosofia naturale"; ed in lingua inglese "philosophy" mantiene l'antico, lato significato) e se le scienze naturali e la filosofia formarono un tutto unico durante la maggior parte della storia dell'antica Grecia.

La filosofia greca volle dare una interpretazione razionale anche agli eventi naturali, sostituendo il *logos* al *mythos;* e la scienza moderna sorgerà, come provincia autonoma della cultura umana, sganciandosi *a sua volta* dal dominio del dogmatismo filosofico del Medio Evo (non dimentichiamo che ancora oggi "teoria" significa etimologicamente, più o meno, "visione divina"). Fino al '700, poi, la parola fisica includerà tutte le scienze della natura (in inglese "medico" si dice ancora *physician)*. Infatti "fisica" corrisponde in greco a "le cose naturali").

Lo strettissimo nesso scienza-filosofia esercitò sulla fisica una duplice influenza: da un lato la filosofia greca fornì la necessaria disposizione all'astrazione sistematica (prima quasi sconosciuta); dall'altro, però (per la erronea svalutazione teorica di ogni lavoro "servile"), impedì alla fisica un dialogo fecondo con la tecnica. Di più, dato che la fisica ebbe la sua esclusiva diffusione attraverso le scuole filosofiche (le uniche organizzazioni culturali esistenti), essa fu ovviamente filtrata da queste scuole, e mescolata a nozioni e principi filosofici. Solo nel III secolo a.C., con Archimede, ci si renderà conto che nessun progresso poteva ottenersi senza la continuità di lavoro di molti (sempre seguendo lo stesso metodo), sì che vedremo Seneca concepire la comprensione del cosmo come un processo "storico", un còmpito senza fine.

I primi inizi di un ragionamento scientifico sistematico si ebbero forse a Mileto, al principio del sesto secolo a.C. Talete, Anassimandro ed Anassimene si posero il problema della sostanza fisica della materia unica primordiale, che doveva stare alla base di ogni ente fisico ed ogni fenomeno: Talete la chiamò "acqua", Anassimene "aria", Anassimandro -- piu cautamente-- preferì non darle alcun nome. Si tentava già, comunque, di spiegare razionalmente un massimo di fenomeni con un minimo di ipotesi. La scuola di Mileto associò inoltre alla "materia primordiale" una legge di conservazione (l'antenata delle molte che costituiscono ora parte essenziale della fisica).

Il risultato più elevato della **scuola di Mileto** fu comunque l'affermare che il moto, «esistente fin dall'eternità», rendeva ragione del principio secondo cui la qualità può venire ridotta a quantità. Qui è adombrato molto dell'essenziale della fisica di tutti i tempi. (Inoltre Talete previde l'eclisse solare del 585 a.C. e scoperse l'elettrizzazione per strofinio.) Anassimandro usò considerazioni logiche di simmetria, applicando il principio di ragion sufficiente ad un caso (di simmetria) in cui interveniva una "condizione di indifferenza": metodo poi usato spesso, da Laplace in poi; e ricorse ad *un modello* meccanico per spiegare il moto apparente del Sole, contribuendo cosi all'enorme rivoluzione del pensiero milese. Nel secolo seguente, Empedocle introdusse delle forze, distinte dalla materia, ed intuì che per un moto equilibrato erano necessari due tipi di forze, attrattive e repulsive (che chiamò poeticamente "amore" e "odio"); con ciò postulando l'esistenza di *cause* nel mondo fisico ed usando il concetto di rapporto causale. Empedocle inoltre "scoperse" che l'aria era un corpo materiale e che la luce si propagava con velocità finita, e concepi la materia come granulare (interpretando le *soluzioni* e le *miscele):* nei pori c'era solo 1' "etere".

Al contrario, Anassagora volle vedere l'aspetto continuo della materia; ed asserì che *l'etere* avesse un moto costante circolare e trascinasse con sé le stelle.

**Nel V secolo a.C.,** il procedere scientifico acquistò molte delle caratteristiche che non perse più: basti dire che il cosmo venne analizzato definitivamente in termini di numeri e misure. «Il Sole è più grande del **Peloponneso»** arriverà a dire **Anassagora, che pure interpretò correttamente l'origine di un meteorite ed attribui alla Luna (considerata con monti e valli e zone abitate!) luce solare riflessa. Anassagora fu, cosi, pioniere della unità dei fenomeni cosmici e terrestri, mettendo sullo stesso piano corpi celesti allora deificati ed oggetti della terra.** Empedocle ed Anassagora, infine, si avvidero anche delle forze centripete.

Più avanti accenneremo ai già tanto noti atomisti: Leucippo, Democrito e, successivamente, Epicuro e Lucrezio.

**Nel IV secolo**, la *scuola pitagorica* scoperse che il creato obbediva a leggi matematiche, e tanto sottolineò l'importanza dell'ordinamento *matematico* per la comprensione della natura, che pose il numero a base essenziale di tutte le cose. **Pitagora**, poi, **affermò la sfericità della Terra, considerata non al centro dell'universo ma una stella come le altre, e -- applicando per la prima volta la matematica ad un fenomeno fisico fondamentale, per di più in un caso (eccezionale per la Grecia) di sperimentazione sistematica-- fece le sue note scoperte sull'armonia musicale**, trovando i rapporti numerici corrispondenti a vari accordi, insieme con i pitagorici di Crotone e Metaponto.

I Pitagorici proiettarono nei cieli i ritrovati della loro teoria dell'armonia musicale, pensando che i pianeti producessero suoni componentisi armonicamente nella «musica delle sfere»: questa fede nell'unità della geometria, della musica e dell'astronomia è interessante --secondo il Sambursky-- perché manifesta fede nell'unità tra uomo e cosmo.

Ancora oggi si è convinti che le leggi fisiche scoperte in laboratorio siano valide pure nell'universo. Teofrasto, discepolo di Aristotele, si porrà con chiarezza, però, il problema, decisamente epistemologico, di *come* potesse il fenomeno *naturale* essere spiegabile col linguaggio *artificiale* della matematica umana.

Eudosso di Cnido (409-356 a.C.) fu il primo a concepire un modello geometrico (a «sfere concentriche») dell'intero cosmo. Eraclide Pontico (388-315 a.C.), fondatore della «teoria degli epicicli» (affermò infatti Venere e Mercurio ruotare attorno al Sole!), fu grande astronomo e pensò -- insieme con alcuni Pitagorici: Iceta, Ecfanto, ecc.—la Terra. dotata di moto rotatorio attorno al proprio asse. **Aristarco da Samo (310-230 a.C.), poi, precorse Copernico di diciotto secoli, esponendo una teoria eliocentrica,** **ed assegnando alla Terra il duplice moto di rotazione e di rivoluzione;** calcolò inoltre le distanze astronomiche ricorrendo a geometria e trigonometria.

Contemporaneamente ad Aristarco visse il grande Archimede, che seppe scoprire le prime leggi fisiche dell'idrostatica, introducendo per di piu il concetto di «peso specifico» (conquista allora difficilissima per la dittatura dell'antitesi aristotelica tra i concetti «assoluti» di *pesante* e *leggero)*. Fama in tutto il mondo gli diedero l'invenzione degli argani e gli studi sulla leva. Le leggi di Archimede, però (come quelle pitagoriche), sono limitate ancora alla sola statica, e non vi entra il tempo. Benché il ruolo del tempo, e precisamente lo stretto nesso tra tempo e moto, verrà presto rivelato dal sommo filosofo e naturalista Aristotele, il quale avverti anche l'importanza della «velocità» in meccanica.

Passiamo a ricordare Eratostene (275-195 a.C.), notissimo per avere calcolato la lunghezza della circonferenza terrestre con straordinaria precisione. Osservazioni ancora più precise, nel campo astronomico, fece Ipparco (circa 190-120 a.C.), che forni --tre secoli prima-- i dati di base per la teoria *tolemaica.* Ipparco catalogò un migliaio di stelle, con le loro coordinate, determinò accuratamente la durata del giorno, e scoperse addirittura il fenomeno della precessione degli equinozi!

Il metodo scientifico pitagorico di esperimentare su diversi strumenti, ed esprimere i risultati in termini generali, tendendo alla formulazione matematica di leggi universalmente applicabili era dunque ottimo. Ma il suo corso fu bloccato da Platone, il quale (non senza le sue ragioni) abbandonò l'idea che la armonia del cosmo potesse essere rivelata attraverso il mondo sensoriale: il creato poteva venir compreso solo dalla matematica pura, mentre il contatto col mondo empirico poteva soltanto oscurare tale comprensione. L'idea che le misurazioni fisiche fossero prive di senso, perché senza nesso con la precisione assoluta dei numeri puri, condusse al disprezzo della sperimentazione quantitativa ed alla pratica rinuncia ad esprimere fatti fisici in termini numerici. Ciò si inseriva nella tendenza dei greci (ma, purtroppo per noi, *non* limitata a quel tempo ed a quel Paese) a sopravvalutare il potere della deduzione, fino a far apparire l'induzione come del tutto superflua. Cosi Aristotele spinse tanto avanti la tendenza a generalizzare, che questa gli divenne di ostacolo.

Un'influenza negativa di Platone, poi, fu quella di sostenne una *differenza* essenziale tra fenomeni celesti e terrestri; quando tale separazione fu accettata e sancita in termini generali e «scientifici» anche da Aristotele, il destino della scienza greca fu segnato: la scienza *moderna* del XVII secolo nacque quando si ridimostrò che le leggi della meccanica terrestre sono invece valide anche per i pianeti. Inoltre Platone --riconosciuto, con Anassagora, che l'armonia dell'universo era dovuta ad una Intelligenza modellatrice-- considerò la causa *teleologica* (finalistica) come l'unica vera spiegazione dei fenomeni cosmici. Non negò altre cause, ma le ritenne prive di importanza per la spiegazione razionale delle cose. Chiamò infine «dottrine del probabile» le scienze naturali.

La teleologia platonica cozzò contro la dottrina atomistica di Democrito; tuttavia la scuola (atomistica) epicurea ebbe molto meno influenza della rivale scuola stoica, fautrice di concezioni finalistiche. Eppure *gli atomisti*: Leucippo (450 a.C. circa), Democrito (circa 460-370 a.C.), Epicuro (341-270 a.C.) e Lucrezio (circa 95-55 a.C.) furono scienziati ricchi di originalità; specialmente Democrito, che è uno dei pensatori piu universali dell' antichità (come riconobbe anche Aristotele). Essi svilupparono un nuovo tipo di ragionamento scientifico, basato sulla prova per *analogia* (ricordiamo quella democritea tra atomi e lettere dell'alfabeto) e *sull'inferenza* dal visibile all'invisibile, con uso di paralleli e «modelli» a guisa d'illustrazione. Ad esempio, Democrito interpretò, cosi, la Via Lattea come costituita da una miriade di stelle. Il razionalismo di Democrito, d'altra parte, lo portò vicino a Platone ed altri predecessori per quanto concerne la coscienza della limitazione dei sensi.

Intorno *all'atomismo* propriamente detto, citiamo solo questo noto passo di Leucippo: «*Questi **atomi**, che nel vuoto infinito sono separati fra loro e che differiscono per forma e per grandezza e per ordine e per posizione, si muovono nel vuoto e, incontrandosi, si urtano: e parte rimbalzano e vengono spinti casualmente, parte invece si collegano a seconda della convenienza di forma, grandezza, ordine e posizione, e restano uniti; e cosi si svolge la generazione di tutto ciò che è cornposto*». Lucrezio esprimerà la chiara idea che un gruppo unico, formato da atomi in moto casuale con urti reciproci, può apparire come un unico corpo in quiete; e, viceversa, descriverà il "moto browniano". Notevoli, ulteriori contributi furono portati da Epicuro, che *compì* il passo dalla teoria atomistica a quella *molecolare*. Inoltre Epicuro ebbe netta la intuizione che «percettibile» ed «impercettibile» sono due «categorie di esistenza» differenti. Infine Leucippo espresse il principio di causalità con le parole: «Nulla si produce a caso, ma tutto con una ragione e necessariamente».

Si diceva di Platone, secondo cui non va fatta confusione tra la «vera causa» (finalistica) e le condizioni che sono necessarie perché essa manifesti sé. Aristotele si impadronì del principio teleologico, lo perfezionò e gli attribuì una importanza eccezionale, secondo l'assioma che tutto ciò che accade si compie per un dato fine, e l'intero universo --con tutto quel che contiene-- è il risultato di un disegno prestabilito. Così Aristotele tese a sistemare le sue scoperte in un modello prefissato, e poi a costruire su di esso una teoria generale che egli proclamava assoluta. Ma, anche nel campo scientifico, la formulazione "dogmatica" è pericolosa, *specie* se anteriore alla raccolta di un numero sufficiente di prove empiriche. Cosicché la influenza aristotelica sulle scienze fisiche fu nel complesso negativa; tanto più che le dottrine fisiche di Aristotele vennero accettate come dogmi per sessanta generazioni. Nessun'altra personalità nella storia della scienza ebbe influenze cosi profonde e durevoli sui pensatori successivi *(culto della personalità).* Il rapido progresso delle scienze naturali negli ultimi secoli è cominciato da quando si è separato (almeno cronologicamente) il momento della ricerca scientifica da quello della ricerca filosofica, cioè da quando gli scienziati lasciarono per lo più ad altri di rispondere alla domanda «a che scopo?», e si limitarono inizialmente a chiedersi «in che modo?» (benché alla lunga la fisica, costruendo modelli interpretativi del mondo via via più evoluti, sia da sempre incamminata a rispondere anche ai vari «perché», in quel processo conoscitivo senza fine che ha luogo per successive approssimazioni -- ciascuna delle quali contiene la precedente come caso particolare).

Il risultato psicologico del modo di pensare post-aristotelico fu che, ogni qual volta esperimento e teoria cozzavano tra loro, venne dato torto all'esperimento (in Italia l'educazione scientifica "moderna" è talmente scarsa, che alcuni giovani studenti pensano tuttora così). Comunque Aristotele fu l'unico scienziato dell'antichità ad accingersi ad una teoria semi-quantitativa della *dinamica* (espressa in forma matematica, e con validità limitata dal confronto con l'esperienza), anche se questa risultò sotto molti aspetti gravemente erronea. Interessante è la dimostrazione della non esistenza del «vuoto», in base a valide considerazioni logiche; e la combinazione di geometria e di materia, ideata dal Nostro per formare il suo concetto di «luogo», non è dissimile dal concetto di *spazio* nella teoria della relatività generale. Effettivamente oggidì il cosmo è concepito in modo ben diverso dalla "scatola vuota" di Newton o degli atomisti greci; e lo «spazio privo di materia» non può dirsi a rigore «vuoto», essendo il "campo metrico" percorso da onde gravitazionali, elettromagnetiche e sede d'azione delle forze nucleari, ecc. Aristotele seppe anche applicare nei ragionamenti il processo di "passaggio al limite". (Isolata restò la sua idea dell'eternità dell'universo in ambedue le direzioni temporali). L'ultima critica ad Aristotele verrà all'inizio del VI secolo d.C. da parte di **Filiponio**, studioso tra l'altro di balistica e meccanica applicata, il quale dimostrò errata la teoria aristotelica dell'impulso (impartito al proiettile dall'aria durante la sua corsa, secondo Aristotele!), basandosi sulle ricerche empiriche iniziate con Archimede e continuate con i primi sviluppi della tecnica di costruzione delle macchine belliche (Filone di Bisanzio, Erone di Alessandria e il romano Vitruvio, ad esempio, possedettero un'autentica mentalità da ingegnere). *Alcuni ragionamenti di Filiponio sono introduttivi alla polemica galileiana.*

Molte cose sarebbero ancora da ricordare, relative agli ultimi secoli della civiltà greca, in particolare riguardo alla *scuola stoica* di Zenone di Cinzia (332-262 a.C.), Crisippo (circa 280-207 a.C.) e **Posidonio (circa 135-51 a.C.)**, nonché riguardo a Seneca (circa 3 a.C. 65 d.C.), Plutarco (circa 46-120 d.C.) ed il già citato Tolomeo (circa 150 d.C.).

Limitiamoci a quanto segue.

Gli *Stoici* - considerato il *continuo,* in tutti i suoi aspetti, la pietra angolare della fisica e postolo come qualità attiva, principio che regge tutti i fenomeni, giunsero ad abbozzare una concezione dei processi termodinamici del mondo inorganico. La stessa "teoria del continuo" stoica, comunque, pur non potendo avvicinarsi alle moderne teorie "del campo unificato" tipo-Einstein, o tipo-Heisenberg, fu ripresa da Newton e dai molti scienziati moderni che sono ricorsi all'"etere". Come si vede, le *idee* scientifiche fondamentali restano le stesse, praticamente, a prescindere dalle risorse tecniche delle rispettive epoche. Inoltre gli Stoici furono i primi a riflettere sulla propagazione delle azioni fisiche in un mezzo continuo, bi- o tri-dimensionale, avente qualità "tensionali". Studiarono i moti ondosi, le onde stazionarie, la propagazione del suono, ecc. **Posidonio** scoperse il *legame* tra la Luna (e persino il Sole) e le maree oceaniche, spiegato con la tensione del «continuo» *(pneuma)* esistente anche tra Luna e Terra. La scuola stoica, poi, approfondì l'esame del nesso tra velocità ed intervalli di tempo, a ciò spinta dallo studio dei famosi paradossi di Zenone di Elea, discepolo di Parmenide, e di un problema analogo posto da Democrito. Crisippo propose audaci soluzioni (simili a quella moderna di introdurre una "lunghezza fondamentale" onde evitare una autoenergia infinita per l'elettrone); e tentò anche di introdurre una terminologia per l'infinitamente piccolo. Notevoli sono i successi stoici nel campo della matematica, tanto da aprire -teoricamente- la strada al calcolo differenziale. Importantissimi i primi barlumi stoici di "dipendenza funzionale", nonché il concetto di "variabile continua" (formatosi all'epoca di Archimede e di Erastotene). Ma tutto ciò non ebbe séguito, e per l'inesistenza della geometria analitica (dovuta alla grave difficoltà greca nel tradurre i ragionamenti nel formalismo matematico), e per l'incapacità di intendere il tempo come variabile indipendente ed i fenomeni come funzione di esso. Infine gli Stoici --come prima i Pitagorici-- ed in particolare Eraclito e, dopo di lui, Empedocle, avanzarono la teoria dei «cicli cosmici», basata sulla stessa logica interna che in tempi molto più prossimi a noi ha portato a collegare fra loro termodinamica, meccanica statistica e concetto di "ritorno all'identico". Precisamente, **Posidonio** scrive: ***«La materia ha una coesione che la tiene insieme e contro la quale il vuoto circostante è impotente. In verità il mondo materiale si conserva mediante una forza immensa, ed a1ternativamente si contrae e si espande nel vuoto seguendo le proprie trasmutazioni fisiche, ora consumato dal fuoco, ora invece dando nuovamente inizio alla creazione del cosmo»***. Ancora una volta è facile constatare come nella storia si ripresentino gli stessi modelli e le stesse associazioni, sia pure in forma nuova ed adeguata allo stadio avanzato raggiunto dalle conoscenze. Interessante è notare che le intuizioni stoiche non potevan essere fondate su alcuna conoscenza delle "leggi dei grandi numeri". Infatti per il disinteresse greco alla *ripetizione* dei fenomeni che già non fossero "naturalmente" periodici (disinteresse che li tenne lontani quasi da ogni esperimentare "artificiale") la scienza greca non ebbe alcuna nozione di statistica. Ad esempio, non venne fatto alcuno studio sui giochi d'azzardo (allora diffusissimi) e sul relativo calcolo delle probabilità. Non è senza significato, secondo Geymonat, che lo studio matematico delle probabilità ed una sperimentazione sistematica siano nati contemporaneamente, nei secoli XVI e XVII.

(Tra parentesi, Ricordiamo anche l'orologio ad acqua e l'organo ad acqua: invenzioni di Otesibio).

**Plutarco, verso il 100 d.C., scrisse la prima opera di "astrofisica" nota**, applicando i metodi e le conclusioni della fisica allo studio della Luna.

Gli argomenti li trasse dalla dinamica (in cui mostra un progredito intuito della gravitazione) e dall'ottica (segnando in certo modo il passaggio dall'ottica geometrica all'ottica fisica, col discutere ad esempio fenomeni di *diffusione*). Influenzato dalla dottrina stoica del "pneuma", Plutarco espose una specie di **teoria generale della gravità**, dicendo che i centri della Terra, del Sole, della Luna erano vari centri di attrazione; e scuotendo la diffusa credenza che l'universo avesse un centro assoluto (fatto incompatibile con la convizione plutarchiana dell'infinità dell'universo). Notevolissimo è il seguente passo: **«*La Luna riceve la garanzia di non cadere proprio dal suo movimento e dallo slancio della sua rivoluzione, esattamente come* i *sassi posti nella fionda non possono cadere per il loro moto circolare vorticoso; infatti ogni cosa è trascinata dal suo semplice moto naturale solo se non è deviata da qualcos'altro. La Luna adunque non è trascinata in basso dal proprio peso, perché la sua tendenza naturale è frustrata dalla sua rivoluzione. Ed anzi, vi sarebbe motivo di meraviglia se essa stesse ferma sempre nel medesimo luogo come la Terra».** A Newton, invece dell'ausilio della famosa mela, sarebbe bastata la lettura di questo brano, eccezionale in ogni sua proposizione!

Interessante è la discussione operata da Seneca intorno alle comete, su cui mostrò di avere idee molto moderne; concludendo però con l'asserire non essere ancora giunto il momento per un giudizio definitivo: «*Accontentiamoci di ciò che abbiamo finora scoperto. Quelli che verranno dopo di noi aggiungeranno il loro contributo alla verità*». Tra le affermazioni apodittiche di Aristotele e le nobili parole di Seneca enorme è il cammino percorso.

A parte consideriamo *l'ottica* greca, che fu dedicata soprattutto a comprendere il meccanismo della visione. La visione è notoriamente un fenomeno complesso, essenzialmente psichico; cosi come la "luce" è una sensazione che si origina (come ovvio) totalmente all'interno degli organismi viventi superiori: in natura non esistono che campi elettromagnetici variabili con diverse frequenze. Una volta che sulla retina si sono formate le immagini, il cervello --riferendosi alla propria esperienza precedente, in particolare dei primi mesi di vita; e precisamente connettendo i dati forniti dagli altri sensi, come il tatto, con dette immagini e con lo stato del cristallino e
l'angolo parallattico dei globi oculari, ricostruisce la forma e la disposizione tridimensionale degli oggetti che ci circondano. Con un processo stupefacente, "proietta" quindi all'esterno di noi --in un certo senso-- la sua ricostruzione, basata sui segnali che gli provengono dai coni e bastoncelli della retina: e *crea* l'effetto della visione. Gli antichi Greci si trovarono disorientati tra la duplice intuizione che qualcosa doveva partire dall'oggetto per recare all'occhio l'informazione, mentre "qualcosaltro" doveva pure uscire da noi "verso gli oggetti".

I Pitagorici abbracciarono l'idea di un'emissione dall'occhio verso l'oggetto. Più concretamente, gli Atomisti si schierarono a favore dell'emissione da parte dei corpi verso l'occhio. Empedocle fu tra i primi a sostenere una combinazione tra i due flussi.

Aristotele pensò ad un collegamento di tipo diverso, senza moto in un senso o nell'altro, ma solo mediante una modificazione del mezzo interposto tra l'occhio e la cosa vista. E aggiunse che se, intorno all'occhio, vi fosse il vuoto completo (cioè l'assenza di ogni "mezzo", *etere* compreso) la visione sarebbe impossibile.

Euclide, il genio matematico, di poco posteriore ad Aristotele, pur accettando la dottrina pitagorica, gettò le basi dell'ottica geometrica tentando di dedurre l'ottica stessa da una dozzina di postulati. I suoi libri, frutto di uno studio sperimentale vasto ma privo di metodo, introducono, ad esempio, il modello di "raggio luminoso" rettilineo, alcuni elementi di prospettiva, le leggi della riflessione e della formazione delle immagini negli specchi piani o sferici, ecc. Un certo progresso venne apportato anche da Eliodoro di Larissa, vissuto in epoca successiva a quella di Tolomeo, e da vari altri.

Si è voluto dare rilievo alla fisica greca, di solito trascurata (poiché spesso nota solo tramite le poco comprensibili e poco competenti traduzioni di –per altro, meritori-- letterati), perché attraverso il suo esame si sono invece potuti toccare alcuni dei temi fondamentali della fisica anche moderna. [Segue una mini-Bibliografia.]

*(Erasmo Recami)*
recami@mi.infn.it